\documentclass[aps,prl,twocolumn,groupedaddress,amsmath,amssymb,showpacs]{revtex4-1}
\usepackage{graphicx}
\usepackage{amsmath}

\begin{document}

\title{General Monogamy Relation between Information-Theoretic Contextuality Inequalities}

\author{Feng Zhu}
\author{Wei Zhang}
\email{zwei@tsinghua.edu.cn}
\author{Yidong Huang}
\affiliation{Tsinghua National Laboratory for Information Science and Technology, Department of Electronic Engineering, Tsinghua University, Beijing, 100084, P. R. China}


\begin{abstract}

We show that the perfect commutation graph is the sufficient tight condition for admitting the noncontextual description of each observable set satisfying it in the yes-no question scenario. With this condition, we propose a method for proving the monogamy relation between two information-theoretic contextuality inequalities by decomposing the total commutation graph into perfect subgraphs. The results offer a powerful tool to investigate the contextuality and to understand quantum information theory. This theoretical work can be experimentally verified in current laboratorial technology.

\end{abstract}



\maketitle


\textit{Introduction.}{---}Contextuality is an important feature of quantum theory, which is different from classical noncontextual hidden variable theory \cite{ks1960}. This counterintuitive property can be indicated by violations of contextuality inequalities \cite{cabello2008_2,kcbs,cabello2008,amselem2009,ysx,amselem2012,kleinmann2012,kaszlikowski2012,chaves_fritz_2012,kaszlikowski2012_2} which can be demonstrated experimentally \cite{cabello_experiment_2009,contexual_experiment_zeilinger,ahrens,duan}. The quantum bounds and non-contextual hidden variable bounds of these inequalities have been investigated by graph-theoretical approach \cite{fang,cabello2010,cabello_explanation,cabello2014,acin2014}. The reason for the difference between quantum bound and classical bound of a contextual inequality is that there exists at least one observable set lacking a joint probability distribution \cite{fine}, where the observable set is required to satisfying the commutation graph \cite{kaszlikowski2012,kaszlikowski2012_2,kaszlikowski2014} which describe the commutation relations among observables.

In the information theory, the Shannon conditional entropy denotes the information needed to describe outcomes of an observable while the other's are given \cite{shannon}. Due to the chain rule, the information-theoretical contextuality inequality (entropic contextuality inequality) can be formulated in classical information theory \cite{kaszlikowski2012,chaves_fritz_2012}. But it can be violated in quantum information theory since the lack of the joint probability distribution, showing the contextuality of quantum theory.

The monogamy relation is the trade-off between the violations of two inequalities. The monogamy relations between two Bell inequalities \cite{verstraete2006prl}, a Bell inequality and a KCBS inequality \cite{kaszlikowski2014}, and two KCBS inequalities \cite{kaszlikowski2012_2} have been demonstrated. They can be used in many fields such as security of quantum key distribution \cite{paw10} and local realism of macroscopic correlations \cite{ramanathan2011}.

As the view of graph-theoretic approach \cite{kaszlikowski2012_2,cabello2014}, the monogamy relation stems from the commutation relations among the observables, where the commutation relations are fundamental in quantum mechanics. The investigation of the monogamy relation between information-theoretic contextuality inequalities contributes to understanding the quantum information theory.



In this Letter, we obtain the condition for existing the noncontextual description of each observable set by investigating the commutation graph in the yes-no question scenario. With this condition, we show a graph-theoretic method to prove the monogamy relation between two information-theoretical contextuality inequalities.



In a commutation graph, each vertex represents an observable and each edge between two vertexes represents that the two observables are compatible. $\{A_1,\ldots,A_n\}$ denotes a set of two-value observables, the value of every observable $A_i$ is $a_i$, $a_{i}{\in}\{1,-1\}$.
${\Pi}_i$ is the projective operator of $A_i$ and $A_i$ can be represented as $A_i=2{\Pi}_i-1$. The widely used yes-no question scenario \cite{kcbs,cabello2008,ysx,kaszlikowski2012,kaszlikowski2012_2,chaves_fritz_2012,contexual_experiment_zeilinger,duan,cabello2010,wright} is the basic scenario since that analysis of contextuality in other scenarios can be expressed in terms of it \cite{peres1991}. In this scenario, two projective operators $\Pi_i$ and $\Pi_j$ are orthogonal if two observables $A_i$ and $A_j$ are compatible. It leads to that the outcomes of $A_i$ and $A_j$ can not be 1 simultaneously.
\begin{eqnarray}
P(A_i=1,A_j=1)=0 \label{exclu}
\end{eqnarray}
Consider that $\{{A_{j_1}},\ldots,{A_{j_m}}\}$ is a subset in which the observables are jointly measurable. The jointly probability distribution of this subset is $P({A_{j_1}}=a_{j_1},\ldots,{A_{j_m}}=a_{j_m})$. By Eq.(\ref{exclu}), we find that this jointly probability distributions of jointly measurable subsets are determinated by $P(A_i=1)$ with $1{\leq}i{\leq}n$, which is the probability that the value of observable $A_i$ is 1.



\textit{Proposition 1.}{---}For an observable set $\{A_1,\ldots,A_n\}$ satisfying a commutation graph $G$, the probability distribution of any jointly measurable subset $P(A_{j_1}=a_{j_1},\ldots,{A_{j_m}}=a_{j_m})$ can be expressed by a linear combination of $P(A_{j_i}=1)$ and a constant if Eq.(\ref{exclu}) holds.

\textit{Proof:} For a joint measurable subset $\{{A_{j_1}},\ldots,{A_{j_m}}\}$, there are $2^m$ possibilities of the set $\{a_{j_1},\ldots,a_{j_m}\}$. They can be divided to three classes according the number of '1' in each possibility.


(a): If more than one elements of $\{a_{j_1},\ldots,a_{j_m}\}$ are equal to 1, for example $a_{j_s}=1$ and $a_{j_t}=1$, the possibility for this case satisfies that $P(A_{j_1}=a_{j_1},\ldots,A_{j_s}=a_{j_{s-1}},A_{j_s}=1,A_{j_s}=a_{j_{s+1}},\ldots,A_{j_{t-1}}=a_{j_{t-1}},A_{j_t}=1,A_{j_{t+1}}=a_{j_{t+1}},\ldots,A_{j_m}=a_{j_m}){\leq}P(A_{j_s}=1,A_{j_t}=1)$ according to the no-disturbance principle \cite{gleason}. Hence, it is equal to zero according to Eq. (\ref{exclu}).
\begin{align}
&P(A_{j_1}=a_{j_1},\ldots,A_{j_{s-1}}=a_{j_{s-1}},A_{j_s}=1,A_{j_{s+1}}=a_{j_{s+1}},\nonumber\\
&\ldots,A_{j_{t-1}}=a_{j_{t-1}},A_{j_t}=1,A_{j_{t+1}}=a_{j_{t+1}},\ldots,A_{j_m}=a_{j_m})\nonumber\\
&=0\label{simpliy_1}
\end{align}



(b): If only one element of $\{a_{j_1},\ldots,a_{j_m}\}$ is equal to 1, for example $a_{j_s}=1$, others are all equal to  $-1$.  According to the no-disturbance principle, $P(A_{j_1}=-1,\ldots,A_{j_{s-1}}=-1,A_{j_{s}}=1,A_{j_{s+1}}=-1,\ldots,A_{j_m}=-1)=P(A_{j_1}=-1,\ldots,A_{j_{s-1}}=-1,A_{j_{s}}=1,A_{j_{s+1}}=-1,\ldots,A_{j_{m-1}}=-1)-P(A_{j_1}=-1,\ldots,A_{j_{s-1}}=-1,A_{j_{s}}=1,A_{j_{s+1}}=1,\ldots,A_{j_{m-1}}=-1,A_{j_{m}}=1)$. While, due to Eq.(\ref{simpliy_1}), $P(A_{j_1}=-1,\ldots,A_{j_{s-1}}=-1,A_{j_{s}}=1,A_{j_{s+1}}=-1,\ldots,A_{j_m}=-1)=P(A_{j_1}=-1,\ldots,A_{j_{s-1}}=-1,A_{j_{s}}=1,A_{j_{s+1}}=-1,\ldots,A_{j_{m-1}}=-1)$. Repeating this processing to all the observables with the values of $-1$, it can be deduced that
\begin{align}
&P(A_{j_1}=-1,\ldots,A_{j_{s-1}}=-1,A_{j_{s}}=1,A_{j_{s+1}}=-1,\ldots,\nonumber\\
&A_{j_m}=-1)=P(A_{j_s}=1)\label{simpliy_2}
\end{align}



(c) If none of $\{a_{j_1},\ldots,a_{j_m}\}$ is equal to 1. With the same deduction above, we can get $P(A_{j_1}=-1,\ldots,A_{j_m}=-1)=P(A_{j_1}=-1,\ldots,A_{j_{m-1}}=-1)-P(A_{j_1}=-1,\ldots,A_{j_{m-1}}=-1,A_{j_{m}}=1)=P(A_{j_1}=-1,\ldots,A_{j_{m-1}}=-1)-P(A_{j_{m}}=1)$. Hence
\begin{align}
P(A_{j_1}=-1,\ldots,A_{j_m}=-1)=1-{\sum_{s=1}^{m}}P(A_{j_s}=1)\label{simpliy_3}
\end{align}


From the Eq.(\ref{simpliy_1})${\thicksim}$(\ref{simpliy_3}), the jointly probability distribution of any jointly measurable subset of any $\{A_1,\ldots,A_n\}$ satisfying the commutation graph $G$ is determined by $P(A_{j_s}=1)$ with $1{\leq}s{\leq}m$. It can be expressed as
\begin{align}
&P(A_{j_1}=a_{j_1},\ldots,A_{j_m}=a_{j_m})\nonumber\\
=&{\prod_{t=1}^m}{\delta_{a_{j_t},-1}}+{\sum_{s=1}^m}\{P(A_{j_s}=1){a_{j_s}}{\prod_{t=1,t{\neq}s}^m}{\delta_{a_{j_t},-1}}\}\label{simpliy_all}
\end{align}
Where $\delta$ is the Kronecker delta function. Hence, it can be concluded that the joint probability distribution $P(A_{j_1}=a_{j_1},\ldots,A_{j_m}=a_{j_m})$ can be expressed as a linear combination of $P(A_{j_s}=1)$, $1{\leq}s{\leq}m$, and a constant.$\hfill$$\Box$



The proposition 1 only uses the condition of Eq.(\ref{exclu}). It can provide a simple way to clarify whether a specific function is the joint probability distribution of an observable set.



\textit{Proposition 2.}{---}$F(A_{1}=a_{1},\ldots,A_{n}=a_{n})$ is a function defined on an observable set $\{A_1,\ldots,A_n\}$ which satisfies a commutation graph $G$ in the yes-no question scenario. $F(A_{1}=a_{1},\ldots,A_{n}=a_{n})$ is the joint probability distribution which recovers the probability distributions of any joint measurable subset $P(A_{j_1}=a_{j_1},\ldots,A_{j_m}=a_{j_m})$ as its marginal distributions if it satisfies that\\
(A). $F(A_{1}=a_{1},\ldots,A_{n}=a_{n}){\geq}0$. \\
(B). ${\sum_{a_i,1{\leq}i{\leq}n}}F(A_{1}=a_{1},\ldots,A_{n}=a_{n})=1$. \\
(C). $F(A_{1}=a_{1},\ldots,A_{n}=a_{n})$=0 when $A_i$ and $A_j$ are compatible and $a_i=a_j=1$.\\
(D). ${\sum_{a_j,j{\neq}i}}F(A_1=a_1,\ldots,A_{i-1}=a_{i-1},A_i=1,A_{i+1}=a_{i+1},\ldots,A_n=a_n)=P(A_i=1)$, where $P(A_i=1)$ with $1{\leq}i{\leq}n$ is the probability that the value of $A_i$ is 1.

\textit{Proof:} Due to the condition (A) and (B), $F(A_{1}=a_{1},\ldots,A_{n}=a_{n})$ is a probability function of $\{A_1,\ldots,A_n\}$. Its marginal distribution is $F(A_{k_1}=a_{k_1},\ldots,A_{k_l}=a_{k_l})={\sum_{a_i,i{\notin}\{k_1,\ldots,k_l\}}}F(A_1=a_1,\ldots,A_n=a_n)$. The condition (C) shows that $F(A_i=1,A_j=1)$, which is the marginal distribution of $F(A_{1}=a_{1},\ldots,A_{n}=a_{n})$, satisfies Eq. (\ref{exclu}) when $A_i$ and $A_j$ are compatible. Hence, $F(A_{1}=a_{1},\ldots,A_{n}=a_{n})$ satisfies the condition of the Proposition 1, by which it can be concluded that the marginal distribution of any joint measurable subset $F(A_{j_1}=a_{j_1},\ldots,A_{j_m}=a_{j_m})$ can be expressed by $F(A_{j_s}=1)$, $1{\leq}s{\leq}m$, with the same formation as Eq. (\ref{simpliy_all}). If condition (D) holds, $F(A_{j_1}=a_{j_1},\ldots,A_{j_m}=a_{j_m})$ is exactly equal to $P(A_{j_1}=a_{j_1},\ldots,A_{j_m}=a_{j_m})$. Hence, $F(A_{1}=a_{1},\ldots,A_{n}=a_{n})$ is the joint probability distribution of $\{A_1,\ldots,A_n\}$.$\hfill$$\Box$



For a commutation graph $G$, the Proposition 2 offers a way to clarify that whether each observable set satisfying $G$ has a joint probability distribution.



\textit{Theorem 1.}{---}For each observable set satisfying a commutation graph $G$ in the yes-no question scenario, there is the joint probability distribution if and only if $G$ is a perfect graph.

\textit{Proof:}


Necessity: If the commutation graph $G$ isn't a perfect graph, $G$ has an odd cycle $C_m$, or an odd cycle's complement $\bar{C}_m$ as its induced subgraph with $m{\geq}5$ \cite{graph_0}. A KCBS-type inequality which is reduced to the Wright-type inequality \cite{wright} in the two values and rank-1 projective operators scenario can be constructed.
\begin{eqnarray}
{\sum_{i,i{\in}G'}}P(A_i=1){\leq}{\alpha}(G') \label{cycle}
\end{eqnarray}
where $G'$ is $C_m$ or $\bar{C}_m$ and ${\alpha}(G')$ is the independent number of $G'$ \cite{lovasz1986}. In quantum theory, there is a specific observable set $\{A_1,\ldots,A_n\}$ and state, under which the left side of Eq.(\ref{cycle}) can reach $\vartheta(G')$ \cite{cabello2014}, where $\vartheta(G')$ is the $Lov\acute{a}sz$ number of $G'$ \cite{lovasz1986,lovasz1979}. According to Ref. \cite{number}, it can be deduced that $\vartheta(C_m)=\frac{m\cos{\frac{\pi}{m}}}{1+\cos{{\frac{\pi}{m}}}}{>}\frac{m-1}{2}={\alpha}(C_m)$ and $\vartheta(\bar{C}_m)=\frac{1+{\cos}\frac{\pi}{m}}{\cos{\frac{\pi}{m}}}{>}2={\alpha}(\bar{C}_m)$ with $m{\geq}5$. In this case, the Eq.(\ref{cycle}) is violated, indicating the lack of a joint probability distribution for the observables of $\{A_{j_1},\ldots,A_{j_m}\}$ satisfying $G'$, which is a subset of $\{A_1,\ldots,A_n\}$. Hence, $\{A_{1},\ldots,A_{n}\}$ does not have a joint probability distribution.



Sufficiency: For an observable set $\{A_1,\ldots,A_n\}$ satisfying the commutation graph $G$, $P(A_i=1)$ is the possibility that the value of the observable $A_i$ is 1. A vector can be constructed as ${\bf p}=(p_1,\ldots,p_n)$, where $p_i=P(A_{i}=1)$ with $1{\leq}i{\leq}n$. The vector ${\bf p}$ satisfies $p_i{\geq}0$ with $1{\leq}i{\leq}n$ and ${\sum_{j_s{\in}C}}p_{j_s}={\sum_{j_s{\in}C}}P(A_{j_s}=1)={\sum_{j_s{\in}C}}P(A_{j_1}=0,\ldots,A_{j_{s-1}}=0,A_{j_{s}}=1,A_{j_{s+1}}=0,\ldots,A_{j_m}=0){\leq}1$, where $C$ consists of $\{A_{j_1},{\ldots},A_{j_m}\}$, representing an arbitrary clique in $G$. The second property is the result of the Proposition 1 and the global exclusivity \cite{cabello_explanation}.

Hence, $\bf p$ is in the fractional vertex packing polytope of $G$, denoted by ${\bf p}{\in}$FVP($G$) (also called QSTAB$(G)$ ) \cite{lovasz1986}. If $G$ is a perfect graph, then FVP($G$)=VP($G$) (or QSTAB$(G)$=STAB$(G)$ ) \cite{lovasz1986}, where VP($G$) (also called STAB$(G)$ ) denotes the vertex packing polytope of $G$ \cite{lovasz1986}. According to the definition of VP($G$), VP($G$)=convex hull $\{ {\bf q}^{(k)}$: ${\bf q}^{(k)}$ is a stable labeling of $G \}$, where $q^{(k)}_i=1$ if vertex $A_i$ is in the $k$-th stable set of $G$, otherwise, $q^{(k)}_i=0$. Hence, there exists a set of $\{{\alpha}_k\}$, where ${\alpha}_k{\geq}0$, $\sum_k{\alpha}_k=1$ and ${\bf p}={\sum_k}{\alpha_k}{\bf q}^{(k)}$.

We can construct a function on $\{A_1,\ldots,A_n\}$
\begin{align}
F(A_1=a_1,\ldots,A_n=a_n)={\sum_k}\{{\alpha_k}{\prod^n_{i=1}}{\delta_{a_i,{2{q^{(k)}_i}-1}}}\} \label{joint_prob}
\end{align}
where ${\delta}$ is the Kronecker delta function. $F(A_{1}=a_{1},\ldots,A_{n}=a_{n})$ satisfies the condition (A) of the Proposition 2. With the equations of $\sum_{a_i}{\delta_{a_i,{2{q^{(k)}_i}-1}}}=1$, $F(A_{1}=a_{1},\ldots,A_{n}=a_{n})$ also satisfies the condition (B) of the Proposition 2 due to ${\sum_{a_i,1{\leq}i{\leq}n}}F(A_{1}=a_{1},\ldots,A_{n}=a_{n})={\sum_{a_i,1{\leq}i{\leq}n}}\{{\sum_k}{\alpha_k}{\prod^n_{i=1}}{\delta_{a_i,{2{q^{(k)}_i}-1}}}\}={\sum_k}{\alpha_k}=1$. According to the definition, for any $k$, $q^{(k)}_i$ and $q^{(k)}_j$ cannot be 1 simultaneously while vertex $A_i$ and $A_j$ are adjacent. Hence, $F(A_{1}=a_{1},\ldots,A_{n}=a_{n})$ satisfies the condition (C) of the Proposition 2. With the equation of ${\delta}_{1,{2{q^{(k)}_i}-1}}={q^{(k)}_i}$, the condition (D) also holds according to that $F(A_i=1)={\sum_{a_j,j{\neq}i}}F(A_1=a_1,\ldots,A_{i-1}=a_{i-1},A_i=1,A_{i+1}=a_{i+1},\ldots,A_n=a_n)={\sum_{a_j,j{\neq}i}}{\sum_k}\{{\alpha_k}{\delta}_{1,{2{q^{(k)}_i}-1}}{\prod^n_{j=1,j{\neq}i}}{\delta}_{a_j,{2{q^{(k)}_j}-1}}\}={\sum_k}{\alpha_k}{\delta}_{1,{2{q^{(k)}_i}-1}}={\sum_k}{\alpha_k}{q^{(k)}_i}=p_i=P(A_i=1)$. Since $F(A_{1}=a_{1},\ldots,A_{n}=a_{n})$ satisfies all the conditions of the Proposition 2, $F(A_{1}=a_{1},\ldots,A_{n}=a_{n})$ is the joint probability distribution of $\{A_1,\ldots,A_n\}$. The analysis can be applied on any observable sets satisfying $G$, demonstrating the necessity of the theorem.$\hfill$$\Box$




The Theorem 1 shows that the perfect commutation graph is the sufficient tight condition for admitting a noncontextual description of the observable set satisfying it in the yes-no question scenario.

\begin{figure}[!htb]
\includegraphics[width=3.5 in]
{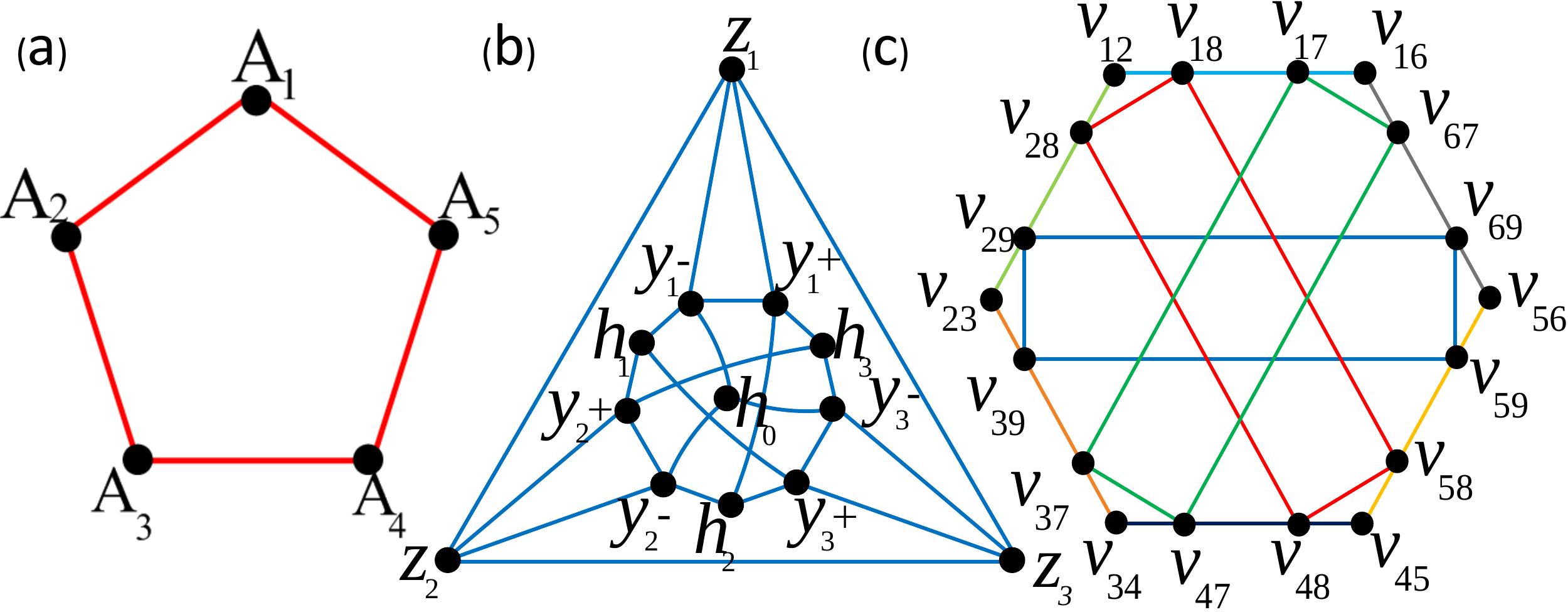}
\caption{Some typical commutation graphs used in investigations of contextuality inequalities: (a)the KCBS inequality\cite{kcbs}. (b)state-independent contextuality with qutrit\cite{ysx}. (c)state-independent contextuality with qudit\cite{cabello2008}. (Observables in the same rectangle or the same side of hexagon are compatible. These relations haven't be indicated for concision.)  \label{fig1}}
\end{figure}

To construct nontrivial contextuality inequalities, the commutation graphs shouldn't be perfect. For example, three typical commutation graphs investigated previously are shown in FIG. \ref{fig1}, while each graph has at least one pentagon as its induced subgraph.

For an odd cycle, which is the simplest imperfect graph, an information-theoretic contextuality inequality can be constructed due to the Theorem 1. Two information-theoretic contextutality inequality can't be violated simultaneously when some constraints are added in the two commutation graphs. We noticed that a method for proving the monogamy relation between two KCBS-type inequalities is demonstrated \cite{kaszlikowski2012_2}. In the method, one needs to decompose the commutation graph into chordal subgraphs which admit noncontextual descriptions. Then, it should be verified that the sum of noncontextual bounds corresponding to the subgraphs is equal to the sum of noncontextual bounds corresponding to original commutation graphs \cite{kaszlikowski2012_2}. Since the set of perfect graphs contains the set of chordal graphs \cite{chordal} and the noncontextual bounds of entropic inequalities are zero \cite{kaszlikowski2012,chaves_fritz_2012}, a similar method can be utilized in the monogamy relation between two information-theoretic contextuality inequalities with less conditions according to the Theorem 1.



\begin{figure}[!htb]
\includegraphics[width=3.6 in]
{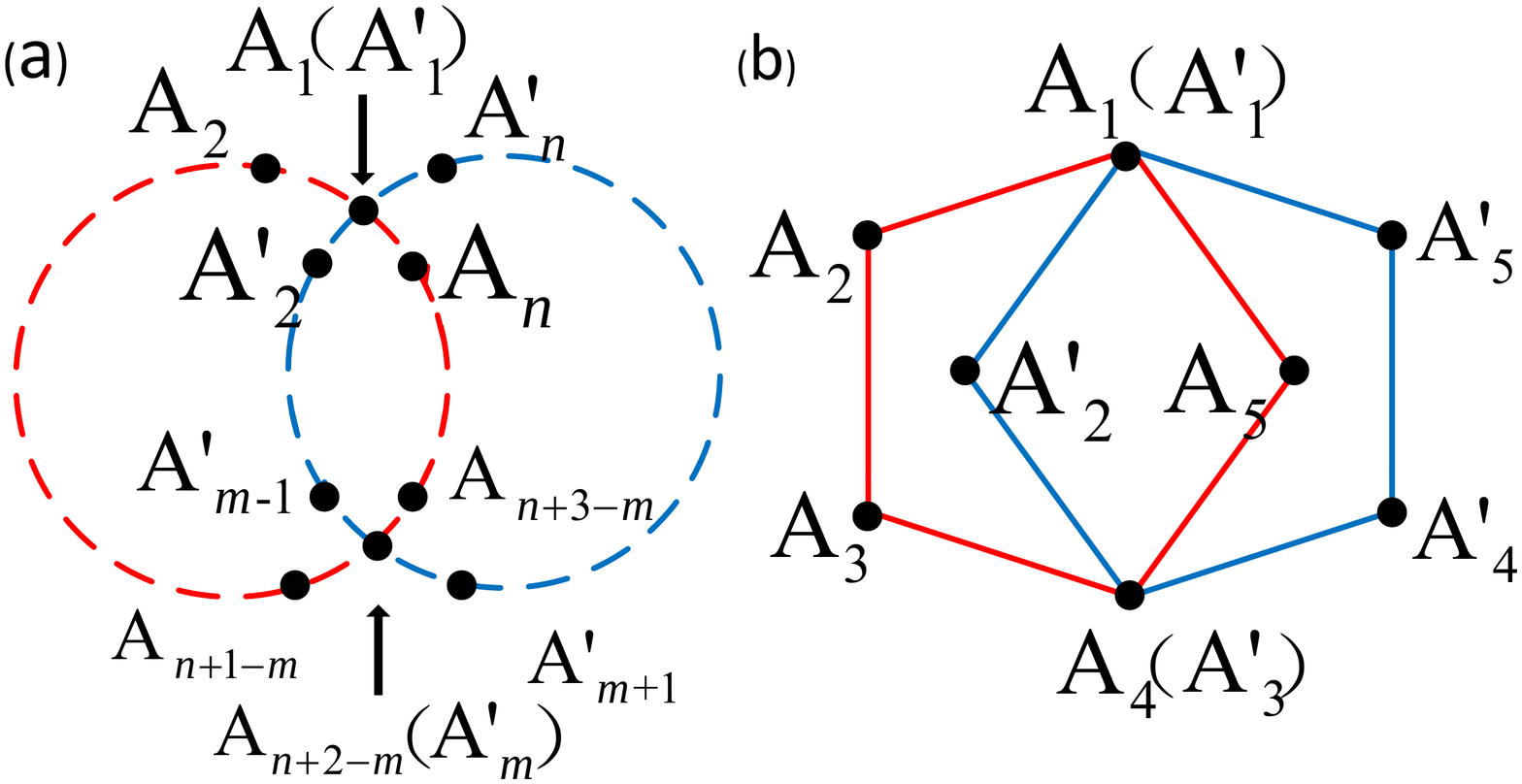}
\caption{The commutation graph of two odd cycles sharing two vertexes.\label{fig}}
\end{figure}

\textit{Theorem 2.}{---}In the yes-no question scenario, two observable sets $\{A_1,{\ldots},A_n\}$ and $\{A'_1,{\ldots},A'_n\}$ with $n{\geq}5$ satisfy the commutation graphs of two odd cycles $C_n$ which share two common vertexes $A_1=A'_1$ and $A_{n+2-m}=A'_m$ as shown in FIG. \ref{fig}(a), where $m$ is an arbitrary number.

(i) There is no monogamy relation between two KCBS-type contextuality inequalities as shown in Eqs.(\ref{t2-1}).

\begin{subequations} \label{t2-1}
\begin{align}
{\sum\limits^{n}_{i=1}}P(A_i=1){\leq}\alpha(C_n)\label{t2-1:a}\\
{\sum\limits^{n}_{i=1}}P(A'_i=1){\leq}\alpha(C_n)\label{t2-1:b}
\end{align}
\end{subequations}

(ii) There is the monogamy relation between two information-theoretic contextuality inequalities as shown in Eqs.(\ref{t2-2}).

\begin{subequations} \label{t2-2}
\begin{align}
-{\sum\limits^{n-1}_{i=1}}H(A_i|A_{i+1})+H(A_1|A_n){\leq}0\label{t2-2:a}\\
-{\sum\limits^{n-1}_{i=1}}H(A'_i|A'_{i+1})+H(A'_1|A'_n){\leq}0\label{t2-2:b}
\end{align}
\end{subequations}

where $H$ denotes the Shannon conditional entropy.


\textit{Proof:} Let us consider the simplest case of $n=5$ as shown in FIG. \ref{fig}(b). Eqs.(\ref{t2-1}) is reduced to Eqs.(\ref{t2-3}).

\begin{subequations} \label{t2-3}
\begin{align}
{\sum\limits^5_{i=1}}P(A_i=1){\leq}\alpha(C_5)\label{t2-3:a}\\
{\sum\limits^5_{i=1}}P(A'_i=1){\leq}\alpha(C_5)\label{t2-3:b}
\end{align}
\end{subequations}

From the proof of sufficiency in the Theorem 1, there is a state $|\varphi{\rangle}$ and an observable set $\{A_1,{\ldots},A_5\}$ in which $A_i=2|v_i{\rangle}{\langle}v_i|-1$ satisfying the odd cycle commutation graph $C_n$, where $\sum^5_{i=1}P(A_i=1)=\sum^5_{i=1}{|\langle{v_i}|\varphi\rangle|^2}>\alpha(C_5)$ leading the violation of Eq.(\ref{t2-3:a}). One can construct a set of $A'_i$, in which $A'_1$ and $A'_3$ are $A_1$ and $A_4$, respectively, while other observables are defined by $A'_{7-i}=2|v'_{7-i}{\rangle}{\langle}v'_{7-i}|-1$, $i\in\{2,3,5\}$, in which $|v'_{7-i}{\rangle}=\cos{\kappa}|v_i{\rangle}+\sin{\kappa}|\phi_0{\rangle}$,  $0{<}\kappa{<}\arccos{\sqrt{\frac{\alpha(C_5)}{\sum^5_{i=1}{|\langle{v_i}|\varphi\rangle|^2}}}}$ and $|\phi_0{\rangle}$ is orthogonal to $|\varphi{\rangle}$, $|v_1{\rangle}$, $|v_2{\rangle}$, $|v_3{\rangle}$, $|v_4{\rangle}$ and $|v_5{\rangle}$. It can be shown that Eq.(\ref{t2-3:b}) is violated since $\sum^5_{i=1}P(A'_i=1)=\sum^5_{i=1}{|\langle{v'_i}|\varphi\rangle|^2}{>}\sum^5_{i=1}{{\cos}^2{\kappa}|\langle{v_i}|\varphi\rangle|^2}{>}\alpha(C_5)$. As a result, the Eqs.(\ref{t2-2}) can be violated simultaneously by the state $|\varphi{\rangle}$ and two specific observable sets $\{A_1,\ldots,A_5\}$ and $\{A'_1,\ldots,A'_5\}$ satisfying the commutation graph shown in FIG. \ref{fig}(b). The proof of (i) can be generalized to the cases of $n>5$ with the similar method.

On the other hand, Eqs.(\ref{t2-2}) is reduced to Eqs.(\ref{t2-4}) in the case of $n=5$.

\begin{subequations} \label{t2-4}
\begin{align}
-{\sum\limits^4_{i=1}}H(A_i|A_{i+1})+H(A_1|A_5){\leq}0\label{t2-4:a}\\
-{\sum\limits^4_{i=1}}H(A'_i|A'_{i+1})+H(A'_1|A'_5){\leq}0\label{t2-4:b}
\end{align}
\end{subequations}

For any observable set satisfying the commutation graph shown by FIG. \ref{fig}(b), $\{A_1,A_2,A_3,A_4,A'_4,A'_5\}$ constructs an even cycle $C_6$, and $\{A_1,A'_2,A_4,A_5\}$ constructs an even cycle $C_4$. They are all perfect graph \cite{graph}. From the Theorem 1, both of them have joint probability distributions in the yes-no question scenario. Hence, both of their information-theoretic contextuality inequalities hold according to Ref. \cite{kaszlikowski2012} and Ref. \cite{chaves_fritz_2012}.

\begin{align*}
-H(A_1|A_2)-H(A_2|A_3)-H(A_3|A_4)-H(A_4|A'_4)&\\
-H(A'_4|A'_5)+H(A_1|A'_5)&{\leq}0\\
-H(A_1|A'_2)-H(A'_2|A_4)-H(A_4|A_5)+H(A_1|A_5)&{\leq}0
\end{align*}

The sum of equations above shows the monogamy relation between Eq.(\ref{t2-4:a}) and Eq.(\ref{t2-4:b}). For the case of $n>5$, the commutation graph shown in FIG. \ref{fig}(a) also can be decomposed into two even cycles of $\{A_1,A_2,\ldots,A_{n+1-m},A_{n+2-m},A'_{m+1},\ldots,,A'_n\}$ and $\{A_1,A'_2,\ldots,A'_{m-1},A_{n+2-m},A_{n+3-m},\ldots,A_n\}$. The proof of (ii) above can be generalized to the cases of $n>5$ with the similar method.$\hfill$$\Box$




Here we show a method for proving the monogamy relations between two information-theoretic contextuality inequalities in the Theorem 2. The key is the decomposition of the total commutation graph into two perfect subgraphs.

The monogamy relation between two information-theoretic contextuality inequalities is able to be demonstrated experimentally. For instance, the observable set shown in FIG. \ref{fig}(b) could be constructed in a qu-dit system, such as an optical system encoded in polarization and path of a photon \cite{amselem2009,amselem2009}.



{\it Conclusion.}{---} In the two propositions of this paper, we show that single observable marginal probabilities $P(A_i=1)$ with $1{\leq}i{\leq}n$ uniquely determinate all probability distributions of jointly measurable observables in the yes-no question scenario. It can be used to verify the existence of the joint probability distribution $F(A_1=a_1,\ldots,A_n=a_n)$ which can recover $P(A_i=1)$ with $1{\leq}i{\leq}n$ as marginal probabilities.

Based on the two propositions, we prove the Theorem 1 that the commutation graph $G$ is a perfect graph is the necessary and sufficient condition for the existence of the joint probability distribution for each observable set satisfying $G$ in the yes-no question scenario. It is proved by constructing a nontrivial contextuality inequality when the commutation graph $G$ isn't perfect and finding out the joint probability distribution $F(A_1=a_1,\ldots,A_n=a_n)$ while $G$ is perfect, respectively. This result provides a powerful tool to determine whether an observable set exists a noncontextual description and formulate its joint probability distribution, which is the core of investigations of quantum contextuality.

According to the Theorem 1, we investigate the monogamy relation between two information-theoretic contextuality inequalities by decomposing the commutation graph into perfect subgraphs. We show that there is the monogamy relation between two information-theoretic contextuality inequalities, in which the observable sets satisfy two odd cycles shared two observables. But there is no monogamy relation between two KCBS-type contextualilty inequalities satisfying the same commutation graph. It reveals some interesting characteristics of conditional entropy, which may contribute to investigations of differences between classical information theory and quantum information theory.



{\it Acknowledgement.}{---}This work was supported by 973 Programs of China under Contract No. 2011CBA00303 and 2013CB328700, Basic Research Foundation of Tsinghua National Laboratory for Information Science and Technology (TNList).




\end{document}